\documentclass[12pt,preprint]{aastex}
\usepackage{natbib,epsfig}

\begin{document}

\newcommand{\hii}{{\rm H}{\sc ii}}
\newcommand{\uchii}{{\rm UCH}{\sc ii}}
\newcommand{\iras}{{\it IRAS}}
\newcommand{\htwo}{${\rm H_2}$}
\newcommand{\nhone}{NH$_3$(1,1)}
\newcommand{\nhtwo}{NH$_3$(2,2)}
\newcommand{\nhthree}{NH$_3$(3,3)}
\newcommand{\ammonia}{NH$_3$}
\newcommand{\methanol}{CH$_3$OH}
\newcommand{\htwoo}{H$_2$O}
\newcommand{\hsixalpha}{H66$\alpha$}
\newcommand{\thcoone}{$^{13}$CO ${(J=1\rightarrow0)}$}
\newcommand{\ceioone}{C$^{18}$O ${(J=1\rightarrow0)}$}
\newcommand{\ceio}{C$^{18}$O}
\newcommand{\cseo}{C$^{17}$O}
\newcommand{\nh}{NH$_3$}
\newcommand{\um}{$\mu$m}
\newcommand{\percc}{cm$^{-3}$}
\newcommand{\persqcm}{cm$^{-2}$}
\newcommand{\emunits}{pc~cm$^{-6}$}
\newcommand{\kms}{km~s$^{-1}$}
\newcommand{\kmsperpc}{km~s~$^{-1}$~pc~$^{-1}$}
\newcommand{\vlsr}{${\rm v}_{\rm lsr}$}
\newcommand{\msun}{M$_\odot$}
\newcommand{\lsun}{L$_\odot$}
\newcommand{\jyperbeam}{Jy beam$^{-1}$}
\newcommand{\mjyperbeam}{mJy beam$^{-1}$}
\newcommand{\jyperbeamkms}{Jy beam$^{-1}$km~s$^{-1}$}
\newcommand{\mjyperbeamkms}{mJy beam$^{-1}$km~s$^{-1}$}
\newcommand{\arcseconds}{$''$}
\newcommand{\degrees}{$^\circ$}

\title{An Infalling Torus of Molecular Gas Around the Ultra-Compact \hii\ Region G28.20-0.05}

\author{Peter K. Sollins, Qizhou Zhang, Eric Keto, Paul T. P Ho}
\affil{Harvard-Smithsonian Center for Astrophysics, 60 Garden Street, Cambridge, MA, 02138, psollins@cfa.harvard.edu}

\begin{abstract}

We present new observations of the ultra-compact \hii\ region
G28.20-0.05 in 23~GHz continuum, and the \nhone, \nhtwo, and \nhthree\
lines. To explain the complicated kinematics of the molecular gas, we
propose a model consisting of two components. One component is an
infalling, equatorial torus of molecular gas, whose dense central
region has been ionized to form the ultra-compact \hii\ region. The
second component is a larger expanding molecular shell driven by some
type of wide-angle outflow or wind. We estimate that the infall
component includes more than 18~\msun\ of molecular gas. We calculate
the central mass to be 79~\msun, probably comprising more than one
star. The arrangement of the molecular material suggests a connection
to the other disk-like structures seen around massive young stars. The
central star in this case is more massive, and the whole region may be
more evolved than other similar objects such as \iras\ 20126+4104,
\iras\ 18089-1732, G192.16-3.84, and AFGL5142.

\end{abstract}


\section{Introduction} \label{sec:intro}

Much research has been done attempting to extend the now well-defined
process of low-mass-star formation to intermediate- and high-mass-star
formation. The accepted standard model for low-mass-star formation, as
described in \citet{shu87}, involves a molecular cloud core which
collapses to form a protostar with an accretion disk. A bipolar
outflow forms. Accretion through the disk slows. Eventually, the cloud
is dispersed and the protostar quasi-statically contracts onto the
main sequence, radiating away its gravitational energy. \citet{kah74},
however, pointed out that, because of their enormous luminosity, the
Kelvin-Helmholtz time-scale for contraction of a massive star is much
shorter than for low mass stars. Massive stars may radiate away their
gravitational energy so quickly that they approach the main sequence
and become very luminous before the surrounding protostellar envelope
has finished collapsing \citep{gar99}. \citet{kah74} and \citet{wol87}
calculated that in this case the pressure of the stellar radiation may
be adequate to stop accretion under certain
circumstances. Observationally, however, several examples are now
known of stars or protostars with masses of order 10~\msun\ or less,
i.e. early B-type, which appear to be undergoing accretion through a
disk \citep{zha98b,zha02,she99,chi04,beu04a}. Most of these sources
are known to also include an outflow. Thus, some aspects of the highly
successful standard model of low-mass-star formation have been
observed in stars up to stellar masses of 10~\msun, despite the
potential radiation pressure problem.

At the high end of the mass scale, a very different type of accretion
scenario has been proposed to explain observations of the \uchii\
region G10.6-0.4. The ionized gas in G10.6 has a radius of about
6000~AU (0.03~pc), and contains a grouping of massive young stars,
with a luminosity of $9.2 \times 10^5$\lsun, a mass of 150~\msun,
emitting $2.2 \times 10^{50}$ Lyman continuum photons per second,
including at least one early O-type star \citep{sol05a}. Outside the
\uchii\ region, molecular gas can be seen to be infalling and rotating
in absorption in the \nhthree\ line \citep{ho86,ket87b,ket88}. This
molecular accretion flow proceeds all the way up to the ionization
front \citep{sol05a}. In the ionized gas, \hsixalpha\ observations
show that the ionized gas is also moving inward and settling into an
ionized disk in the center of the \uchii\ region
\citep{ket05}. \citet{ket02b} proposed a model in which the ionized
gas is gravitationally trapped by the mass of the central stars,
allowing the molecular accretion flow to pass through a stalled
ionization front and continue as an ionized accretion flow. The
particular combination of mass and luminosity of the stellar cluster,
and angular momentum in the molecular gas puts G10.6 in a regime in
which the thermal pressure of the ionized gas is confined by the
gravitational force of the central stars, and the molecular gas does
not settle into a disk, but rather a rotating, somewhat flattened,
quasi-spherical accretion flow. This is a very different scenario than
the low-mass standard model in which molecular gas accretes
quasi-statically through a rotationally supported accretion disk.

In order to investigate a regime of massive-star formation in between
the grouping of O-stars in G10.6 and the individual early B-type stars
seen to have disks and outflows, we have observed G28.20-0.05. G28 is
an \uchii\ region at a distance of $5.7 \, ^{+0.5}/_{-0.8}$~kpc
\citep{fis03}, with a radius of about 3400~AU (0.017~pc). Using the
formula of \citet{cas86}, the far infrared luminosity based on IRAS
fluxes is $1.6\times 10^5$~\lsun, corresponding to one O8V star, which
would have a mass of 31~\msun, and an ionizing flux of $7.4\times
10^{48}$ Lyman continuum photons per second\citep{vac96}. Of course,
if more than one star is present, the luminosity does not uniquely
determine the mass or flux of ionizing photons. Using a broader range
of infrared measurements, \citet{wal03} get $1.8\times 10^5$~\lsun,
consistent with the \citet{cas86} value when scaled to our assumed
distance. When compared to G10.6, G28 has a lower luminosity, smaller
radius, and presumably different central mass. These factors should
put G28 into a different accretion regime than the previous work
described above.

\section{Observations} \label{sec:obs}

We observed the \uchii\ region G28.20-0.05 with the NRAO Very Large
Array (VLA)\footnote{The National Radio Astronomy Observatory is a
facility of the National Science Foundation operated under cooperative
agreement by Associated Universities, Inc.} on four occasions in 2003,
with the phase center at $\alpha(2000)=\rm{18^h42^m58^s.10}, \,
\delta(2000)=-4^o 13'57''.87$. We observed the (1,1), (2,2), and (3,3)
inversion lines of \ammonia\ in the D-configuration, yielding a
resolution of about $3''$ in natural weighted maps. We also observed
the (2,2) and (3,3) lines in the hybrid BnA configuration yielding a
resolution of about $0''.3$ in uniform weighted maps.

We observed 3C286 to set the absolute flux scale, and 1849+005 to
calibrate the phases. The passband response was calibrated using 3C273
for the (1,1) and (2,2) D array data, and 3C454.3 for the rest of the
data. After initial flux and phase calibration, all visibilities were
also self-calibrated for phase and amplitude, using as the model a map
made from the velocity-integrated visibilities. The resolutions and
noise levels, as well as some physical parameters of the lines
observed, in each of the maps are listed in Table \ref{tab:data}.

As in earlier work on the inversion lines of \ammonia\ as seen around
\uchii\ regions \citep{ket88,sol05a}, we make this point regarding
spatial resolution, and the detectability of the line in either
emission or absorption. We achieve a typical sensitivity in a single
channel of the line data of 4~\mjyperbeam. At a resolution of $3''$,
this flux sensitivity means that the $3\sigma$ detection limit in
terms of temperature is about 4~K. So thermal emission of any
appreciable optical depth from a hot molecular core whose actual
temperature is around 100~K should be easily detectable. For the
higher resolution data, the flux sensitivity is similar, but the
synthesized beam is a factor of 10 smaller, resulting in a factor of
100 worse temperature sensitivity. No thermal emission, at any optical
depth, will be detectable from a 100~K hot molecular core when the
$3\sigma$ detection limit is 400~K, as in the $0''.3$ resolution
data. The sensitivity to absorption however, is quite high in the high
resolution maps. Our continuum map, based on the BnA array (2,2) data,
has a peak brightness temperature of 7500~K. The $1\sigma$ noise level
in the line channel maps is 130~K, so very optically thick absorption
could be detected at signal to noise ratios exceeding 50. We thus
expect to detect both line absorption and emission in the D array
data, but only absorption in the BnA array data.

\section{Results} \label{sec:results}

\subsection{The High Resolution Data} \label{sec:hires}

\begin{figure}
\epsscale{0.7}
\plotone{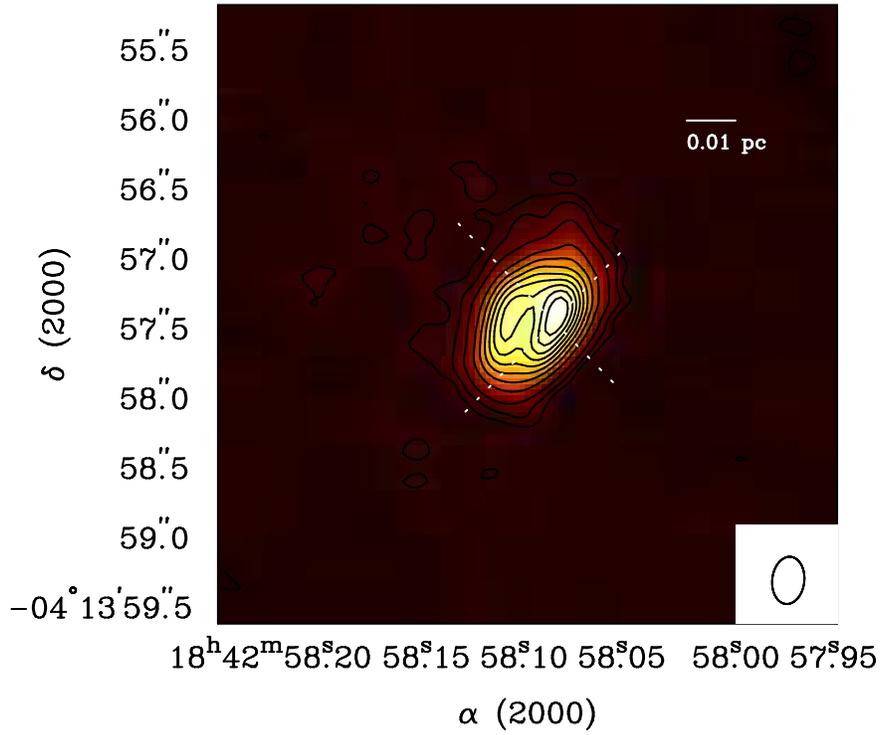}
\caption{The 23~GHz continuum image of G28.20-0.05. The contours are
at 2.5, 5, 10, 20, 30, 40, 50, 60, 70, 80, and 90\% of the peak,
181~\mjyperbeam. The color-scale is linear from -10 to
181~\mjyperbeam. The dotted white lines indicate the position-velocity
cuts shown in Figures \ref{fig:lineposvel}, \ref{fig:aptauposvel}, and
\ref{fig:ratiotauposvel}. The synthesized beam is shown in the lower
right.}
\label{fig:cont}
\end{figure}

The continuum map derived from the BnA array \nhtwo\ observations is
shown in Figure \ref{fig:cont}.  The source is resolved and there are
two peaks separated by $0''.4$, just over one synthesized beam. At the
50\% contour, the continuum emission is $0''.6 \times 0''.8$. The
total flux in the central continuum source is 0.98~Jy. The peak is
181~\mjyperbeam, corresponding to a brightness temperature of
5700~K. Assuming the electron temperature is 10000~K, the peak optical
depth is $\tau_\nu=0.84$, and the peak emission measure is $2.0 \times
10^9$~\emunits. Assuming uniform density and a recombination
coefficient of $3 \times 10^{-13} {\rm cm^3 s^{-1}}$ \citep{ket03},
the electron density is $2.1 \times 10^6$~\percc, the mass of ionized
gas is 0.54~\msun, and the flux of ionizing photons necessary to
balance recombinations in the \uchii\ region is $1.6\times 10^{49}$
per second. If from a single star, this ionizing flux would correspond
to a O6.5 V star of mass 41~\msun\ and luminosity $3.1 \times
10^5$~\lsun. Since this does not match the known luminosity, we
suspect that the central object is a multiple star system, with no
stellar member as early as O6.5. It should also be noted that there
are several assumptions in this calculation (temperature of the
ionized gas, path length associated with the observed continuum flux,
uniform density structure) which could change the resulting spectral
type.

\begin{figure}
\epsscale{1.0}
\plotone{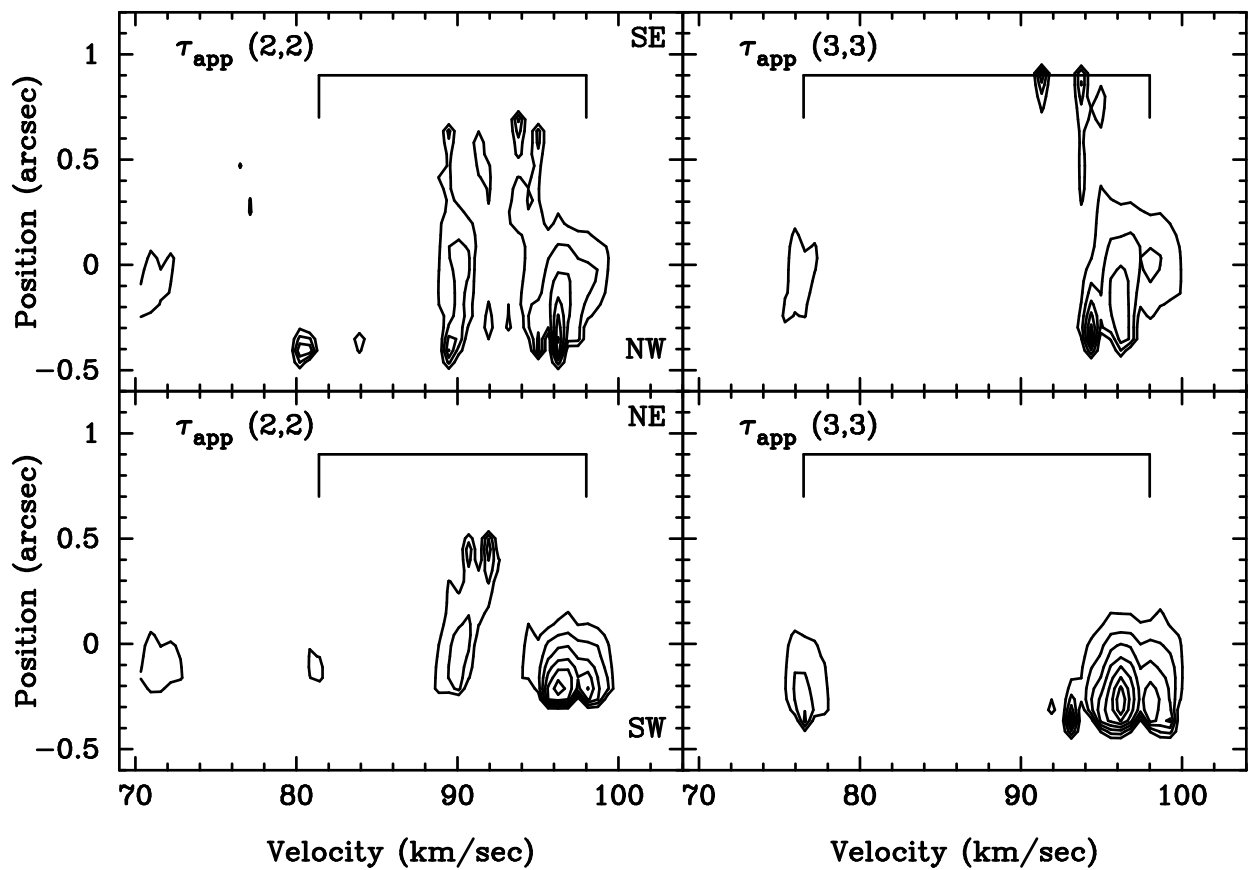}
\caption{Position velocity diagrams in the apparent optical depth of
the \nhtwo\ and \nhthree\ lines. The cuts are NW-SE in the top row,
and NE-SW in the bottom row. The position origin is at the continuum
peak. The optical depth contours are multiples of 0.1. The (2,2) line
clearly shows the two different components, while the (3,3) line does
not show the 90~\kms\ component. The separations of the main hyperfine
component and the inner satellite component are shown for both lines.}
\label{fig:aptauposvel}
\end{figure}

\begin{figure}
\epsscale{1.0}
\plotone{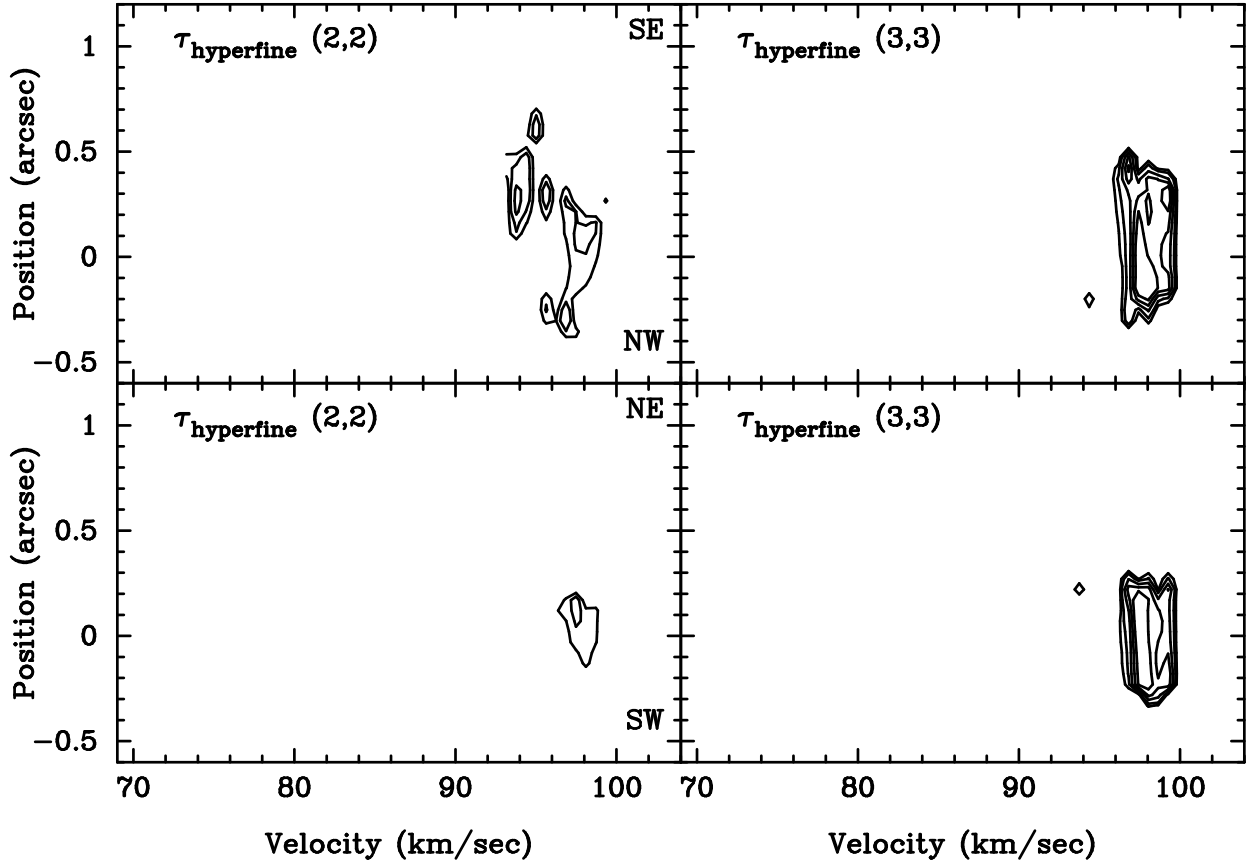}
\caption{Position velocity diagrams in the hyperfine optical depth of
the \nhtwo\ and \nhthree\ lines. The cuts are NW-SE in the top row,
and NE-SW in the bottom row. The position origin is at the continuum
peak. The optical depth contours are multiples of 4. The measurement
of hyperfine optical depth requires the detection of the satellite
hyperfine component. Hence lower optical depth material, detectable in
apparent optical depth, or in the raw absorption line, is suppressed
here.}
\label{fig:ratiotauposvel}
\end{figure}

The key feature of our BnA array line data is that it divides neatly
into two distinct components, one centered at 90~\kms\ which we
associate with an outflow or expansion, and one centered at 97~\kms\
which we associate with infall. The two components are clearly
separated in the \nhtwo\ position-velocity diagrams in Figures
\ref{fig:lineposvel}, \ref{fig:aptauposvel} and
\ref{fig:ratiotauposvel}. The lower resolution data fix the ambient
velocity of the cloud at 95.5~\kms, so one component is blue-shifted,
while the other is red-shifted, hence the infall and outflow
interpretation. Figures \ref{fig:momentpanels} and \ref{fig:90mom1}
also help to distinguish the two components, and will be discussed in
detail below.

\subsubsection{The 90~\kms\ Outflow Component} \label{sec:90}

The outflow component is cool and optically thin. It is detected over
the entire face of the continuum source, and its mean velocity varies
little as a function of position. The first and second moments of the
90~\kms\ component can be seen in the first two panels in Figure
\ref{fig:momentpanels}. The moments of the 90~\kms\ component are
integrated from 84~\kms\ to 92~\kms. Compared to the 97~\kms\
component, both moments of the 90~\kms\ component vary little with
position. But as seen in Figure \ref{fig:90mom1}, the first moment of
the 90~\kms\ component does have a clear pattern. The velocity pattern
shows parallel stripes monotonically increasing from 89.7~\kms\ on one
side to 91.4~\kms\ on the other side. The optical depth of the line in
the 90~\kms\ component is relatively low. There is no detectable
absorption in the first satellite hyperfine line corresponding to the
90~\kms\ component, meaning that the optical depth in the main line is
everywhere less than 2.7. Because the 90~\kms\ component is detected
in (2,2) but not (3,3) we can put an upper limit on its temperature of
30~K. It is important to note that the 90~\kms\ component extends over
the entire face of the \uchii\ region.

\subsubsection{The 97~\kms\ Infall Component} \label{sec:97}

The rest of the panels in Figure \ref{fig:momentpanels} show the
kinematics of the 97~\kms\ component. The most prominent feature in
Figure \ref{fig:momentpanels} is the sharp NW-SE line in the BnA array
maps of the first and second moments. This is the edge of the 97~\kms\
component. Because the 90~\kms\ component extends over the entire face
of the continuum emission, we know the sharp line must be due to a cut
off in the absorption from the 97~\kms\ component. In the direction
perpendicular to the edge, the beam is $> 0''.2$, so the sharpness of
the edge is striking. This edge suggests the presence of a disk-like
structure, perhaps a toroid, surrounding the \uchii\ region. If the
toroid were inclined, it would naturally produce absorption over only
half of the face of the \uchii\ region.

\begin{figure}
\epsscale{1.0}
\plotone{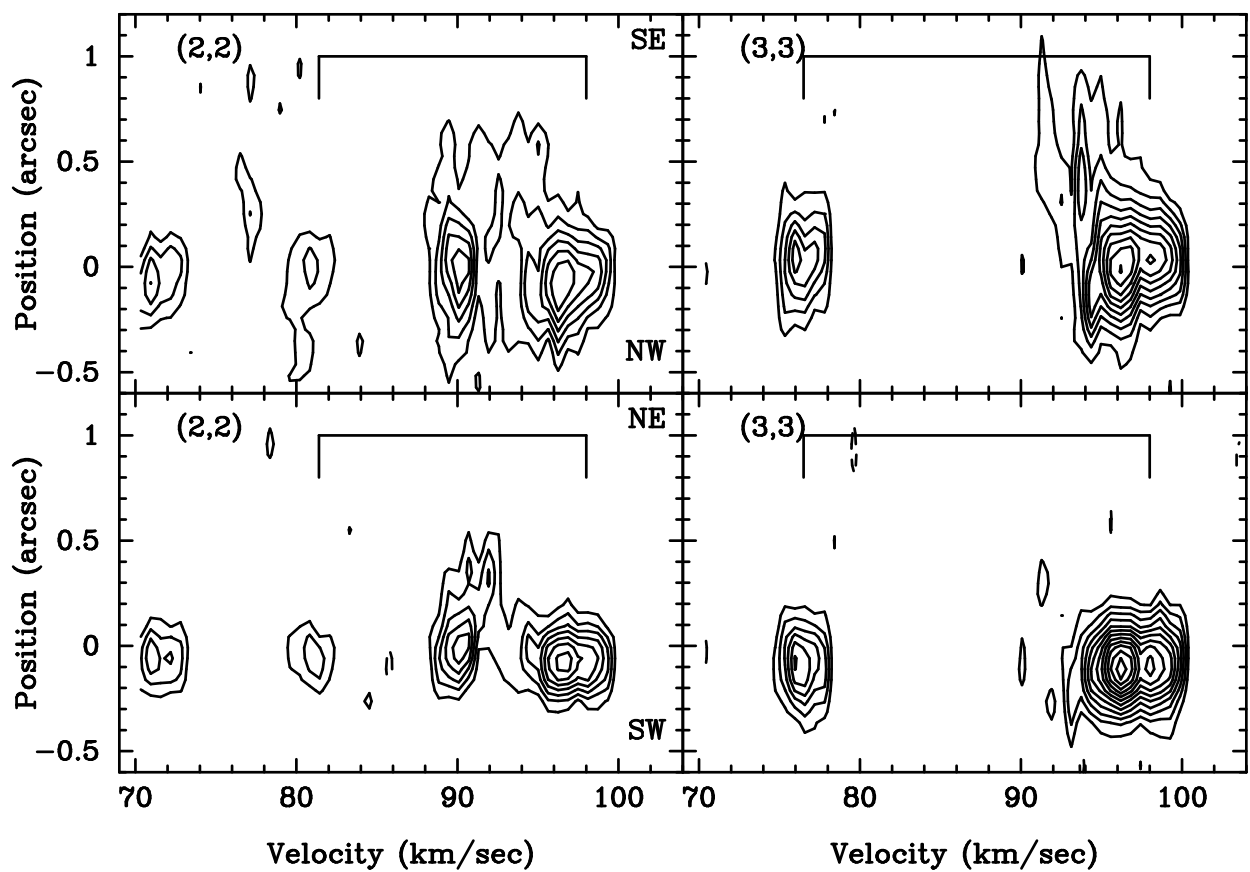}
\caption{Position velocity diagrams in the \nhtwo\ and \nhthree\
absorption lines. The cuts are NW-SE in the top row, and NE-SW in the
bottom row. The position origin is at the continuum peak. The contours
are multiples of 10~\mjyperbeam. Higher contours indicate deeper
absorption, with solid contours used instead of dashed lines for
clarity. The (2,2) line clearly shows the two different components,
while the (3,3) line does not show the 90~\kms\ component. The
separations of the main hyperfine component and the inner satellite
component are shown for both lines.}
\label{fig:lineposvel}
\end{figure}

The spatial pattern of the mean velocity of the 97~\kms\ component
shows infall and possibly weak rotation. The pattern, from northwest
to southeast goes from 95~\kms\ to 97~\kms and back down to
94~\kms. This is reminiscent of the ``off-center bulls-eye''
characteristic of combined infall and rotation discussed in
\citet{ket88} and observed in G10.6-0.4
\citep{ket87b,ket02a,sol05a}. If rotation is responsible for the
offset of the position of fastest infall, then the axis of rotation
must be northeast-southwest, with the projection into the plane of the
sky of the angular momentum vector pointing
northeast. Position-velocity cuts in both the direction of the axis of
rotation (NE-SW) and perpendicular to the axis of rotation (NW-SE) are
shown in Figure \ref{fig:lineposvel}. The most red-shifted infalling
gas can be seen at 100~\kms. While the offset of the position of
fastest infall in the first moment map is conspicuous, it is difficult
to see any additional evidence of rotation in the position-velocity
diagram.

The infalling gas is warm and very optically thick. Figures
\ref{fig:aptauposvel} and \ref{fig:ratiotauposvel} show the same
position-velocity cuts as Figure \ref{fig:lineposvel}, but in the
apparent optical depths, and the hyperfine optical depths instead of
in the actual line absorption. (For an explanation of apparent and
hyperfine optical depths, see the Appendix.) The separations between
the main hyperfine component and first satellite hyperfine component
are 16.6~\kms\ and 21.5~\kms\ for the (2,2) and (3,3) lines
respectively, and in the (2,2) the outer satellite line is at a
separation of 25.8~\kms. Only the most optically thick gas could ever
be detected in the much weaker satellite lines. It is the red-shifted,
infalling gas which is detectable in the satellite lines, and is the
most optically thick. Since infall models generally predict strong
central condensation \citep{shu77,ter84}, we expect the most
red-shifted infalling gas to be the densest, and most optically thick,
and it is, with peak hyperfine optical depths of 16 in the (2,2) and
47 in the the (3,3). The (2,2) and (3,3) peak optical depths
correspond to a rotational temperature of 280~K, much warmer than the
30~K upper limit for the outflow component, which is detected only in
(2,2). Using the 280~K temperature and an abundance of \ammonia\
relative to \htwo\ of $10^{-7}$, we calculate the mass of the gas
detected in the infall component to be 9~\msun, using either the (3,3)
or the (2,2) optical depth, i.e. the two lines are consistent. Note
that this is only a fraction of the mass of the entire infall
component, since only that part of the infall component which lies in
front of the continuum source can be observed. Thus mass of the infall
component should be at least 18~\msun, or perhaps more. The largest
source of uncertainty here is the abundance, which may be uncertain by
as much as an order of magnitude.

\subsection{The Low Resolution Data} \label{sec:lowres}

The molecular core surrounding the \uchii\ is detected in all three
lines we observed in the low angular resolution mode and shows typical
evidence of internal heating. Figure \ref{fig:lowresmom0} shows 3~$''$
resolution, velocity-integrated maps in all three lines. The (1,1) map
includes the most extended emission, while the (3,3) emission is the
most compact. This is consistent with central heating of the molecular
core from the \uchii\ region, similar to the molecular gas in the
regions surrounding massive protostars, such as AFGL5142
\citep{zha02}. All three line maps show two peaks immediately next to
the \uchii\ region, the brighter to the northeast, the fainter to the
southwest. We use these peaks to define an axis for a
position-velocity cut running from 30\degrees\ west of south, to 30
east of north. The more extended emission runs northwest to southeast,
so we use that to define another position-velocity cut. Both cuts are
indicated in Figure \ref{fig:lowresmom0}. The absorption toward the
\uchii\ region can be seen in (2,2) and (3,3) as negative contours at
the center of the zeroth moment maps.

The position-velocity cut in the (1,1) data establishes the velocity
of the ambient cloud. Figure \ref{fig:lowresposvel} shows the
position-velocity cuts indicated in Figure \ref{fig:lowresmom0}. The
first satellite hyperfine components of the \nhone\ line are separated
from the main line by $\pm$~7.8~\kms, and the main line has an
intrinsic line-strength only 3.6 times that of the inner
satellites. Since the kinematics in this object involve relative
velocities greater than 8~\kms, and since the satellite lines can be
seen strongly in emission, the position-velocity diagrams for the
(1,1) line are confusing with strong blending of the main line and the
the inner satellites. Near the \uchii\ region, it is impossible to
tell where emission and absorption from the different hyperfine
components is blending. However, even in the (1,1) position-velocity
diagrams, away from the \uchii\ region, one can see the gas returning
to its ambient velocity, around 95.5~\kms, shifting to about 97~\kms
in the far southeast of the cloud.

In the (2,2) and (3,3) position-velocity diagrams, the kinematics are
less confused, showing the same outflow component detected at high
angular resolution. Off the \uchii\ region, most of the emission is
near the ambient velocity, 95.5~\kms. At the position of the \uchii\
region we see several components in the main line, separated in
velocity. There is blue-shifted absorption in both (2,2) and (3,3)
detectable from 88 to 93~\kms. There is also high velocity red-shifted
emission out to 105~\kms. The absorption is the same outflow component
detected in the high resolution data, and the emission is the back
side of the outflow, which could not be detected at high
resolution. At the position of the \uchii\ region, the apparent
optical depths at 90~\kms\ in (2,2) and (3,3) are 0.25 and 0.05
respectively, giving a rotational temperature of 20~K where we have
assumed the filling factors are equal to one for both lines. This is
consistent with the upper limit of 30~K for the temperature of the
outflow component, derived from the high resolution data. The fact
that the back side of the outflow is only detected at the position of
the \uchii\ region implies that it is no larger than the synthesized
beam. The outflow cannot be too much smaller than the beam, however,
since the emission from the back side of the outflow is strong,
reaching a brightness temperature of 16~K in the (3,3) line, since
emission much smaller than the beam will be strongly beam diluted.

The low angular resolution data also give a hint of the existence of
the infall component, but only in the satellite lines, and mainly in
(3,3), emphasizing the high optical depth and temperature of the
infall component. The satellite is only barely detected in absorption
in (2,2), but in (3,3) it is strongly detected. Since the intrinsic
line-strength of the inner satellite in the (3,3) line is only about
3\% of the main line, detecting the inner satellite strongly means the
optical depth of the gas must be very high. The satellite absorption
is at 75 to 78~\kms. Since the (3,3) satellite line is offset by
21.5~\kms\ from the main line, the corresponding high-optical-depth
absorbing material should appear in the main line from 96.5 to
99.5~\kms, exactly the velocities at which we detect the infall
component at higher resolution (see Figure \ref{fig:ratiotauposvel}
especially). But if the optical depth toward the \uchii\ region is so
high from 95 to 100~\kms, why does the position-velocity diagram show
little absorption and even some emission in that velocity range at the
position of the continuum source? This apparent contradiction does not
mean that the satellite line is somehow wrong in predicting the
presence of optically thick main-line absorption. Instead, in this low
spatial resolution data where the synthesized beam is larger than the
continuum source, line emission in the range 95 to 100~\kms\ from the
area near the \uchii\ region can fill in the absorption which would
otherwise be dominant.

In the low resolution data there are two clumps detected in all three
lines, one northeast, the other southwest of the \uchii\ region. The
satellite lines are detectable in these two clumps, only in the (1,1)
line. Following the method of \citet{ho83} for determining a rotational
temperature when optical depth is calculable for only one rotational
state, we estimate the temperature of both clumps to be 30~K. Based on
the \nhone\ emission, the column densities of \ammonia\ are $1.2
\times 10^{15}$\persqcm\ and $1.0 \times 10^{15}$\persqcm\ in the
northeast and southwest clumps respectively. Assuming an \ammonia\
abundance relative to \htwo\ of $10^{-7}$ \citep{van98}, the northeast
clump contains 12~\msun\ of gas, and the southeast clump 10~\msun.

\section{Discussion} \label{sec:discussion}

\subsection{The Model}  \label{sec:model}

There are six key observational results that any model of this source
must include. First, there must be two components, an infall component
and an outflow component. Second, the outflow component must be seen
in absorption over the entire face of the \uchii\ region, while the
infall component must be seen over only half, cutting off in a sharp
northwest-southeast line. Third, while the velocities of the infall
component show a large projection effect over the $1''$ continuum
source, the velocities of the outflow component do not project out
nearly as strongly, that is the velocities do not vary strongly with
position, and do not return to the ambient velocity at the edge of the
continuum source. Fourth, the infall component must have high optical
depth and a warm temperature (280~K), while the outflow component must
have low optical depth and a lower temperature (20 to 30~K). Fifth,
the outflow component must be smaller than the D array beam since the
emission from the back side of the outflow is seen in the low
resolution data only at the position of the \uchii\ region. At the
same time, the outflow component cannot be much smaller than the D
array beam since the emission is not too strongly beam diluted. Sixth,
the model should explain the elongated shape of the continuum source,
parallel to the sharp edge of the infall component. The model that we
describe here fits all of these observed results.

Figure \ref{fig:cartoon} shows our preferred model for G28. The infall
component and the central continuum source are an inclined, infalling,
possibly slowly rotating toroid, whose central region has been ionized
by the central star out to a radius of 3400~AU (0.017~pc). The toroid
is similar to those proposed in \citet{bel05}. The ionized gas may
resemble the photo-ionized disks modeled in \citet{hol94} except that
in this case there is no rotationally supported disk, just the
infalling toroid. The outflow component is a molecular shell swept up
by a larger, more tenuous expanding bubble of ionized gas. The radius
of the shell is roughly 8300~AU (0.04~pc).

Our proposed model fits all the key observational results. The
infalling and slowly rotating torus can be seen against part of the
continuum source, but not all of it. The rear side of the torus
obviously cannot be seen in absorption. The torus is undetectable in
emission in the D-array data because of its small size compared to the
low resolution beam. The outflow component provides the 88-92~\kms\
absorption, and the 100+~\kms\ emission seen at low resolution. The
outflow is larger, more fully filling the beam at low resolution, and
thus is detectable in emission. Because the expanding shell is large
compared to the continuum source, its line-of-sight velocity does not
return to the ambient velocity at the edge of the continuum
source. The projection which is at work on the velocities of the
infall component therefore does little to change the apparent velocity
of the expanding shell. Because the infall component is closer to the
central source it should be warmer than the outflow, which is mostly
likely an isothermally shocked molecular shell. The infall component
will naturally be more optically thick, since infalling gas is
naturally centrally condensed. If the continuum emission is from
photo-ionization of the densest central region of the toroid, that
continuum emission region ought to be intrinsically flattened. Thus
the inclination needed to get the sharp edge in the first moment map
of the infall component would also naturally provide the elongation of
the continuum source. In this way, the proposed model fits all of our
key observational results.

The infall component is consistent with the existence of a central
source having a mass of 79~\msun, while the infalling gas itself may
include an amount of gas similar in mass. The local velocity is
95.5~\kms, and the most red-shifted part of the infalling gas is at
100~\kms. We will assume that this fastest infalling gas is at the
radius of the continuum source, which has a deconvolved radius of
$0.''6$, or 3400~AU (0.017~pc). If the continuum source is
intrinsically circular, then the elongation implies an inclination of
45\degrees. Assuming the gas is in free-fall toward the central
source, and all the velocity is in the plane of the disk-like
structure, our estimate of the central mass, given by

\begin{equation}
M = \frac{R (v_{in}-v_0)^2}{2G sin^2 i}
\end{equation}

is 79~\msun. This is much larger than, but consistent with the lower
limits of 41~\msun, based on the ionizing flux, and 31~\msun, based on
the far-IR luminosity. (If the velocity is not purely in the plane of
the disk-like structure, then the maximum velocity seen might be
entirely along the line of sight, in which case $sin i = 1$ and the
derived central mass would be 40~\msun.) The luminosity also sets an
upper limit of 31~\msun\ on the largest star in the region, so this is
likely to be a multiple system. This is not surprising given the high
multiplicity of early type stars \citep{pre99}. The mass seen in
absorption in the infall component itself, 9~\msun, could be only a
small fraction of the mass of the toroid depending on the geometry. At
the very least, the total mass is more than 18~\msun, since in
absorption we only detect the near side. This is reminiscent of
disk-like structures seen around very young early B and late O type
stars which have masses which are an appreciable fraction of the
stellar mass \citep{beu04a,zha02,zha98a}.

The outflowing molecular gas could be part of a jet-driven bipolar
outflow as seen around protostars, or it could be the product of
pressure driven expansion of the \hii\ region, or even a spherical
wind. Following \citet{ket02a}, we calculate the Bondi radius of the
ionized gas. Given the mass as derived from the infall, the radius at
which gravity and thermal pressure of the ionized gas balance is
240~AU, where we have assumed an ionized gas temperature of
10000~K. The radius of the detected continuum emission is 3400~AU
(0.017~pc), so one could reasonably expect that the ionized gas is in
a phase of pressure driven expansion. This also justifies our earlier
assumption of constant density in the ionized gas. However, the
velocity of the outflow component is smaller than what one would
expect for pressure driven expansion. The absorption and emission seen
in the low resolution (2,2) data are separated by a maximum of 17~\kms
(88~\kms\ to 105~\kms), so the expansion velocity is 8.5~\kms. As a
sound speed, this corresponds to a temperature of 3200 Kelvin for the
ionized gas. Since the ionized gas is certainly hotter than that, the
expansion may be impeded by the ram pressure of an infalling
envelope. Still, the gravity of the central star should be weak
compared to the force of the pressure imbalance between the ionized
gas and the molecular gas. The ionized gas should be undergoing
dynamic pressure driven expansion. Given a diameter of $3''$, equal to
the size of the D array synthesized beam, the expansion speed implies
a dynamic age of 5200 years. But if the \uchii\ region had an earlier
gravitationally confined stage as described in \citet{ket02a}, the age
of the source could be much larger. Another possible explanation for
the expansion velocity being smaller than the sound speed of the
ionized gas is that the outflow is not undergoing purely spherical
expansion, but is rather a more directed bipolar outflow. In that
case, the outflowing molecular gas might not be moving just along the
line of sight, but also with a velocity in the plane of the sky.

The two clumps, one to the northeast, the other to the southwest, seen
in Figure \ref{fig:lowresmom0}, do not fit clearly into our model. The
two clumps are separated from the \uchii\ region by roughly 0.2~pc
(41,000~AU). Thus, on the scale of Figure \ref{fig:cartoon}, they are
off the page. The question of how to connect the gas in those clumps
to the gas seen in the high resolution data remains, however. First we
should note that if there were no high resolution data available, we
would probably have interpreted the clumps as evidence of a rotating
disk. The \nhthree\ emission in Figure \ref{fig:lowresmom0} is
elongated along the northeast-southwest cut, and shows a velocity
shift from one side of the \uchii\ region to the other side. This is
similar to, although not quite as compelling as, the evidence for a
disk in \iras\ 20126+4104 \citep{zha98b}. The point is moot, however,
since the high resolution data firmly establish the axis of symmetry
as northeast-southwest, and the plane of the toroid as
northwest-southeast. Since the toroid is seen in the high resolution
data in absorption against the southwest side of the \uchii\ region,
we would expect a bipolar outflow to be blue-shifted in the northeast,
and red-shifted in the southwest. This is indeed the sense of the
roughly 1~\kms\ velocity shift observed in the two clumps, as seen in
Figure \ref{fig:lowresposvel}. But the 22~\msun\ observed in the two
clumps is a great deal of mass for the outflow to accelerate. If the
low velocity of the clumps is just a projection effect, and we believe
that the entire 22~\msun\ is expanding with the same three dimensional
speed as the outflow component seen at high resolution, i.e. 8.5~\kms,
the total momentum of the outflow would be
190~\msun~\kms. Alternatively, we could associate the two clumps with
the infall component instead. In that case, the blue-shifted clump in
the northeast would be on the far side, and the red-shifted clump in
the southwest would be infalling on the near side. But the elongation
is perpendicular to the elongation we expect from the geometry of the
torus seen at high resolution. There are several ways to interpret the
two clumps, none without problems. It should be remembered, however,
that these clumps are beyond the physical size of the model, and need
not detract from the fact that the model explains all the other
observed results.

We have considered and ruled out another possible model, that there
are two physically distinct continuum sources, each with its own
associated molecular gas, which, in projection, overlap on the
sky. This model is suggested by several facts. First, the continuum
source is slightly resolved into two peaks. Second, the two components
of the molecular material are quite distinct. Third, it quite
naturally accounts for the sharp edge in the 97~\kms\ as an edge in
the continuum, not in the absorbing material. In this two component
model, the rear continuum source is absorbed by the 97~\kms\
component, and a second, nearer continuum source is absorbed only by
the 90~\kms\ component. This way the 90~\kms\ gas is the nearest to
the observer and can be seen in absorption against all the
continuum. However, we rule out this model for the following
reason. The 90~\kms\ component is seen in the low resolution data to
be paired with the 100+~\kms\ emission. The two must be physically
connected. This would imply that the 100+~\kms\ emission is associated
with the nearer continuum source. But if that were so, it would appear
in absorption against the rear continuum source, which is not
observed. For this reason, we reject this two component model.

\subsection{Comparisons to Other Objects}

We contrast these observations to similar observations of G10.6
\citep{sol05a}. In that case, where accretion is proceeding toward a
central group of massive young stars, there is some flattening
associated with rotation in the molecular gas. But there is no
conspicuous edge like the one we see in G28. In G10.6, the lack of a
sharp edge is cited as evidence that no rotationally supported,
geometrically thin, optically thick disk exists around the stellar
group. While G28 does show such a sharp edge, it also does not appear
to have a rotationally supported disk. The circular rotation velocity
implied by a 79~\msun\ central mass at a radius of 3400~AU with an
inclination of 45\degrees, would be 3.2~\kms, implying a 6.4~\kms\
shift in velocity from one side of the \uchii\ region to the
other. There is in fact less than 1~\kms difference in velocity across
the \uchii\ region. So the sharp edge cannot come from the fact that
the gas has settled into a geometrically thin disk. The flattening
seen could come from an initially flattened geometry of the cloud, as
modeled in \citet{har96}, collapse along magnetic field lines, or
sculpting of the infall component via interaction with the outflow.

There are some similarities and also a key difference between G28 and
other young massive stars observed to have molecular gas in disk-like
structures. \iras\ 20126+4104, \iras\ 18089-1732, and AFGL5142 all
include early B type or late O type stars surrounded by a rotating
disk at roughly 5000~AU scale \citep{zha98b,beu02a,zha02}. These
rotating disks are sub-keplerian, and do not appear to be rotationally
supported, similar to our model of G28. In G192.16-3.82, \citet{she99}
find rotation in water maser spots at radii as small as 1000~AU, also
around an early B type star. In that case, at smaller radii, the
velocity gradient is consistent with Keplerian rotation. In all these
cases, the amount of mass associated with the disk-like structure is
smaller than, but comparable to the estimated mass of the central
object. While G28 has a much more massive central object than any of
these other regions, it seems to have much in common. The toroid
through which infall is proceeding is not rotationally supported at a
radius of 3000~AU. We detect 9~\msun\ of molecular gas in absorption
in the infall component, but that is just in the part of the toroid
that lies directly in front of the continuum source. Certainly there
must be as much mass in the part of the toroid behind the continuum
source as there is in front, and depending on the geometry of the
toroid, the mass might be a factor of a few larger than the measured
9~\msun. Thus the mass of the toroid may be the same order of
magnitude as the central object, similar to the other objects. Unlike
any of these other objects, however, G28 has a strong central
bremsstrahlung continuum source. This difference may be attributable
to the larger mass, and therefore larger ionizing flux of the central
star or stars. G28 could also be a more evolved version of these other
sources, with more time allowing for the evolution of an \uchii\
region. G28 is in some ways similar to \iras\ 20216+4104 and other
similar young massive stars with disks, but the central source in G28
is more massive, and the system may be somewhat more evolved.

\section{Summary}

We propose a qualitative model for the source G28.20-0.05. The model
has two components, an equatorial toroid of infalling molecular gas in
which a central dense region has been photo-ionized to form the
\uchii\ region, and an expanding molecular shell of larger radius,
roughly 8300~AU (0.04~pc). Observationally, the molecular gas is easily
divisible into two kinematically, and physically distinct
components. One is seen in blue-shifted absorption and red-shifted
emission, with low temperature (20-30~K) and relatively low optical
depth which we associate with the outflow. The other is seen only in
red-shifted absorption, with much higher temperature (280~K) and much
higher optical depth (up to more that 45 in the (3,3) line) which we
associate with the infall. We calculate that the central mass
responsible for the infall is 79~\msun, and the that the mass of the
torus is at least 18~\msun. The lack of strong rotation shows that,
although the infalling gas has a flattened geometry, no rotationally
supported accretion disk exists at radii as small as the continuum
source, about 3000~AU.

\section{Appendix: Optical Depths} \label{sec:tau}

We calculate two different types of optical depth in the inversion
lines of \ammonia\ (for an example see Figure
\ref{fig:samplespectra}). The first we call ``apparent optical depth''
and it is found in any given channel by solving
\begin{equation}
T_{line} = T_{cont}(1-e^{-\tau_{app}})
\end{equation}
for $\tau_{app}$, where $T_{cont}$ is the temperature of the continuum
emission and $T_{line}$ is the depth of the line absorption. This
equation assumes no line emission, only absorption. For a given
continuum level, the lowest calculable apparent optical depth is set
by the detection of absorption as different from the continuum level
and the highest calculable apparent optical depth is set by detection
of flux in the channel in question. Apparent optical depth includes a
filling factor, so that a very clumpy gas with locally high actual
optical depth may have a low apparent optical depth if the optically
thick gas does not completely fill the synthesized beam. Apparent
optical depth can be calculated wherever absorption is detected
against the continuum and a lower limit can be derived in any channel
in which the continuum is absorbed down below the detection limit. For
a continuum level of 100 \mjyperbeam\ and noise level in one channel
of a spectrum of 4 \mjyperbeam, if we use $2\sigma$ as our detection
limit, the highest and lowest detectable apparent optical depths are
2.5 and 0.08 respectively. In this case, if the apparent optical depth
were higher than 2.5 it would be indistinguishable from infinity, and
if the apparent optical depth were less than 0.08, the absorption
would be undetectable.

The second type of optical depth we calculate we will call the
``hyperfine optical depth''. This is the optical depth of the main
hyperfine component as calculated from the ratio of the fluxes of a
satellite hyperfine component to the main hyperfine
component. Assuming LTE, and that exactly the same material emits both
the main hyperfine component and the inner satellite, the only
difference in the optical depths of the two lines should be due to
their differing intrinsic line-strengths. Since their optical depths
differ by a known multiplicative constant, $\tau_{sat}=x \tau_{main}$,
the ratio of the fluxes of the two lines can be used to solve for the
optical depth of either line using the following equation.

\begin{equation}
\frac{T_{sat}}{T_{main}}=\frac{1-e^{-\tau_{sat}}}{1-e^{-\tau_{main}}}=\frac{1-e^{-x\tau_{main}}}{1-e^{-\tau_{main}}}
\end{equation}

By convention, we solve for $\tau_{main}$. The details of this
calculation and the hyperfine structure of the different \ammonia\
inversion lines are laid out in \citet{ho83}.  Because only the ratio
of fluxes or temperatures is used, the hyperfine optical depth can be
calculated either when the line is seen in emission, or when the line
is seen in absorption. Also, the use of the ratio means that the
filling factor cancels leaving the hyperfine optical depth a truer
measure of the optical depth. The detectable range of hyperfine
optical depths varies between lines, since the hyperfine structure is
different in different rotational states. For the (2,2) line as seen
in absorption against a continuum brightness of 100 \mjyperbeam, a
noise level of 4 \mjyperbeam\ in one spectral channel, and using a
$2\sigma$ detection limit, the best-case-scenario range of detectable
hyperfine optical depths is 0.5 to 38. For (3,3) the range would be
2.5 to 80. It should be noted that in order to calculate the hyperfine
optical depth both a satellite and the main hyperfine components must
be detected. Since the satellite line has an intrinsically lower
optical depth, it is sensitive to much larger column densities of gas
than the main component. In some cases the satellite may be detectable
in absorption, but the main line may not, for instance if the
continuum source is smaller than the synthesized beam, and widespread,
low-optical-depth emission in the main line fills in the absorption,
as in our lower spatial resolution \nhtwo\ and \nhthree\ data. In this
case, the hyperfine optical depth is not formally calculable. But one
can get an estimate by noting that the apparent optical depth in the
main line should just be the apparent optical depth of the satellite
times the constant $x$. In almost every real circumstance, the
apparent and hyperfine optical depths probe disjoint parts of
parameter space.

\bibliographystyle{apj} \bibliography{bib_entries}

\begin{figure}
%\epsscale{1.0}
%\plotone{f05.eps}
\caption{The first and second moments of the \ammonia\ (2,2) and (3,3)
lines in color and the continuum maps in contours. The contour is 5\%
times the continuum peaks, 188 \mjyperbeam\ in the (2,2) continuum
image, and 305 \mjyperbeam in the (3,3) continuum image, which has a
larger beam. The velocities are indicated by the color bars in
\kms. The first moment is the flux weighted mean velocity of the line,
labeled as ``velocity'', and the second moment is the flux weighted
dispersion, $\sigma$, or 0.425 times the full width of half-maximum
for a Gaussian. For the (2,2) line, moments are first calculated over
the range 84~\kms\ to 92~\kms. These moment maps are labeled
``90~\kms'' and correspond to the 90~\kms\ outflow component. The
moments are then calculated for the (2,2) line over the velocity range
92~\kms\ to 103~\kms. These maps are labeled ``97~\kms'' and
correspond to the infall component near 97~\kms. The moments of the
(3,3) line are calculated over the entire range since the 90~\kms\
component is nearly invisible in (3,3), and only the infall component
is detected. The most striking feature of the data is the sharp line
apparent in the moment maps of the infall component. At positions in
the northeast of the moment maps of the infall component, we see some
leakage of the outflow component into the velocity window used for the
infall component. The synthesized beam is shown in the lower right of
each panel.}
\label{fig:momentpanels}
\end{figure}

\newpage
\addtocounter{figure}{-1}

\begin{figure}
\resizebox{!}{\textheight}{\includegraphics[angle=90]{f05.eps}}
\caption{Continued}
\end{figure}

\begin{figure}
\epsscale{1.0}
\plotone{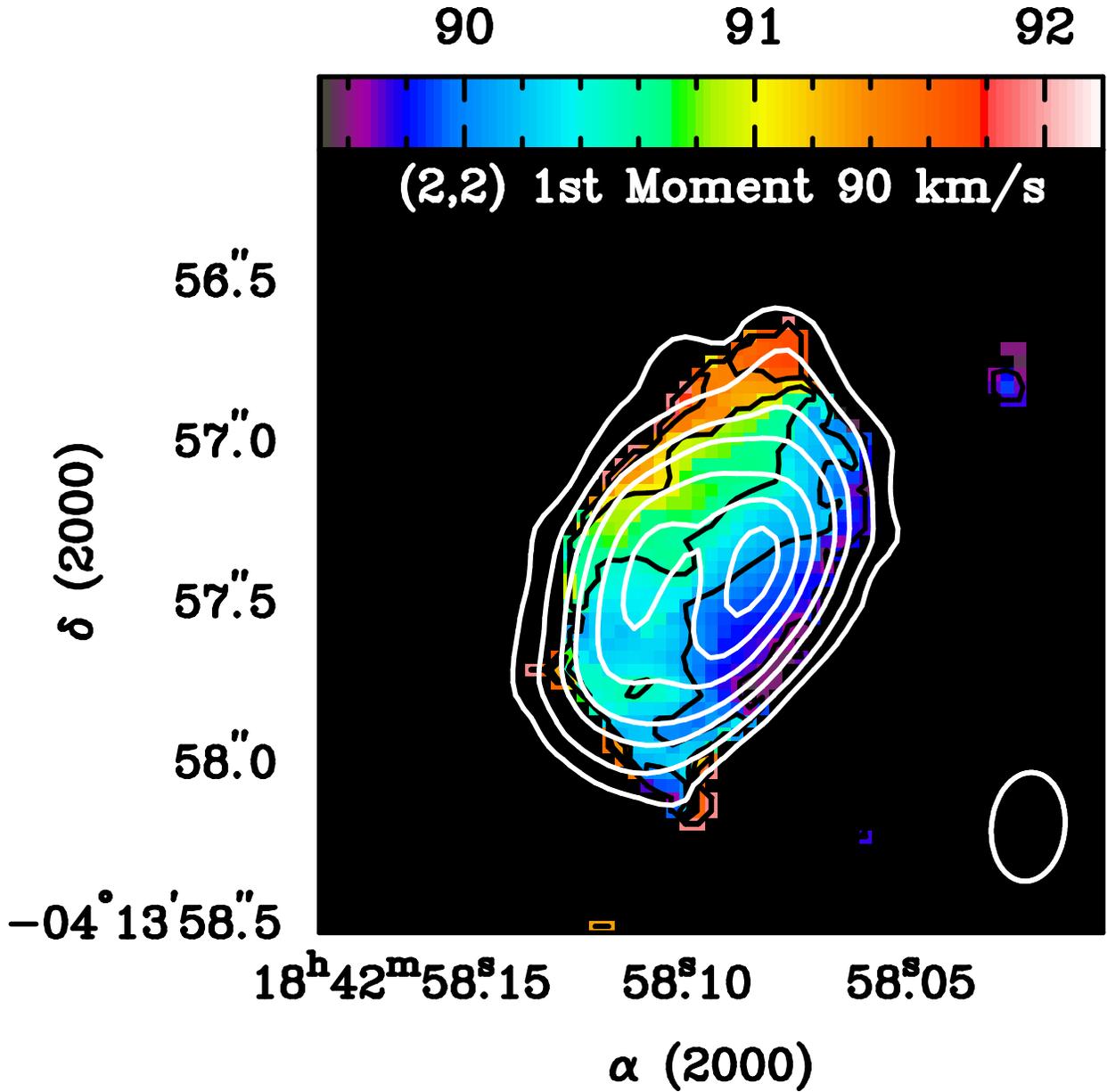}
\caption{First moment of the 90~\kms\ component of the \nhtwo\ line
over the velocity range \vlsr~=~84~-~92~\kms. The colors represent the
mean velocity of the line and run from 89.4 to 91.8. The black
contours also show the velocity field. The white contours are the
continuum map based on the (2,2) observations with contours at 10, 20,
30, 50, 70, 90~\% times 188~\mjyperbeam. The synthesized beam is shown
at the lower right.}
\label{fig:90mom1}
\end{figure}

\newpage

\begin{figure}
\caption{Maps of the velocity integrated line-strength in the (1,1),
(2,2), and (3,3) lines of \ammonia\ at 3$''$ resolution. The
color-scale runs from -0.06 to 0.6~\jyperbeamkms, and the contours are
at 0.05, 0.1, 0.2, 0.3, 0.4, 0.5, and 0.6~\jyperbeamkms. The (1,1),
(2,2) and (3,3) states are 23~K, 65~K, and 125~K above ground
respectively. The coldest gas is seen in \nhone, and is most
extended. The warmest gas is closer to the \uchii\ region. Because
three of the hyperfine components of the (1,1) line are confused, the
line is shown integrated over all three lines, from 80~\kms\ to
100~\kms. The (2,2) and (3,3) lines are integrated from 85~\kms\ to
106~\kms\ so as to include only the main line. The synthesized beam is
indicated in the lower right of each panel. The dotted white lines
indicate the positions of the position-velocity cuts shown in Figure
\ref{fig:lowresposvel}. One cut goes through the more extended gas,
directly northwest to southeast. The other cut goes through the two
emission peaks, 30 degrees west of south to 30 degrees east of north.}
\label{fig:lowresmom0}
\end{figure}

\newpage
\addtocounter{figure}{-1}

\begin{figure}
%\epsscale{0.37}
\resizebox{!}{\textheight}{\plotone{f07.eps}}
\caption{Continued}
\end{figure}

\begin{figure}
\epsscale{1.0}
\plotone{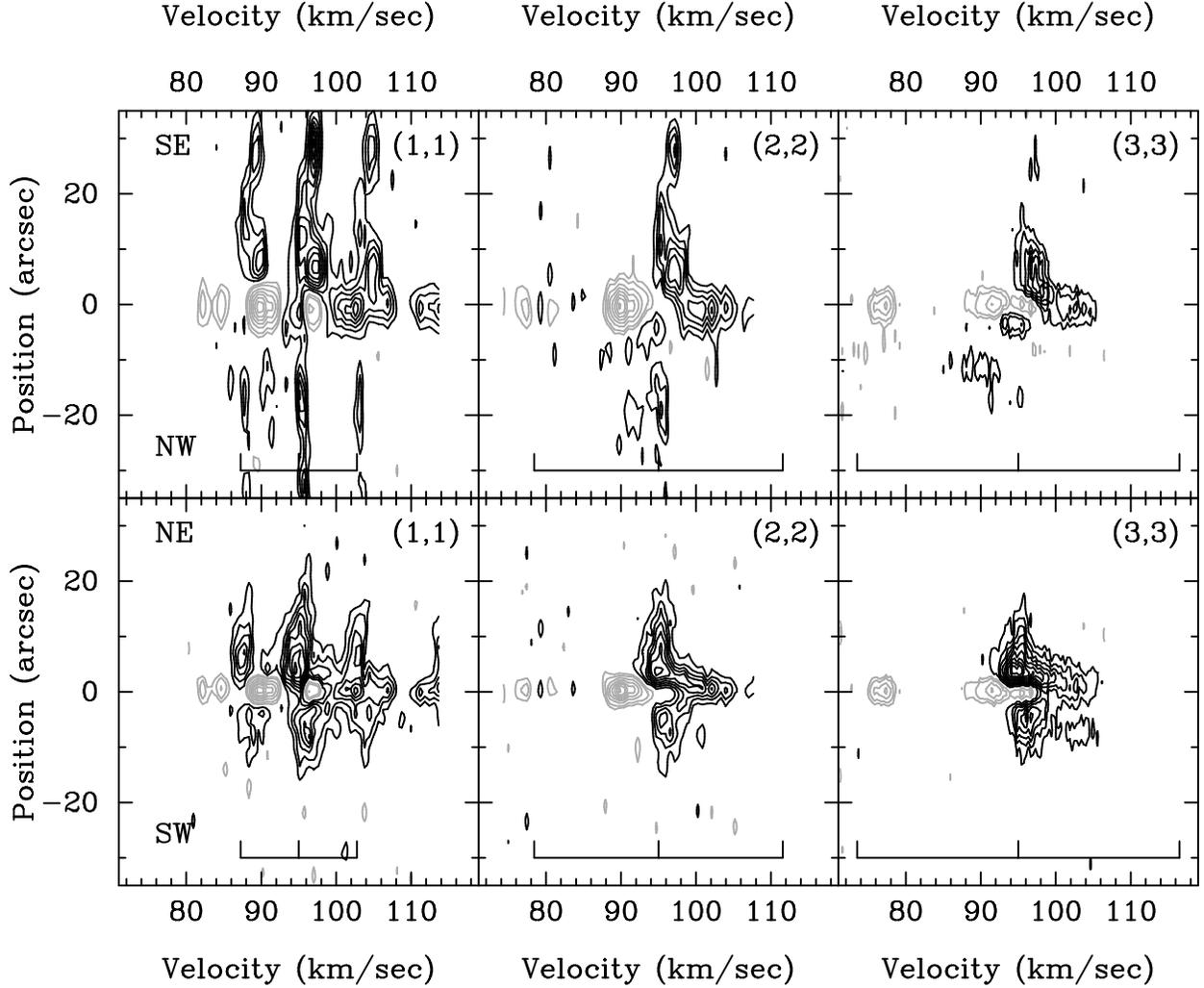}
\caption{Position-velocity diagrams of the (1,1), (2,2), and (3,3)
data at 3\arcseconds\ resolution. The grey contours are -10, -8, -6,
-4, -2, -1 $\times$ 10~\mjyperbeam. The black contours are 1, 2, 3, 4,
5, 6, 7, 8, 9, 10 $\times$ 10~\mjyperbeam. All cuts go through the
\uchii\ region at the origin where absorption is visible against the
continuum. In the top row the cut goes from northwest (negative) to
southeast (positive) and includes the more extended emission. In the
bottom row, the cut goes through the two emission peaks (see Figure
\ref{fig:lowresmom0}) from 30 degrees west of south (negative) to 30
degrees east of north (positive). In all cases, position is positive
to the east and negative to the west so that the position axis
increases with increasing Right Ascension. The position of the two
inner satellite hyperfine components is indicated at the bottom of
each panel relative to 95~\kms. In the (1,1) line, the main line and
both inner satellite lines are included in the bandpass. In the (2,2)
and (3,3) lines, the velocity difference between the main line and the
inner satellite is much larger than for the (1,1) line. Thus, in the
(2,2) and (3,3) data, the spectral window is shifted to lower velocity
in order to include one satellite line in the bandpass.}
\label{fig:lowresposvel}
\end{figure}

\begin{figure}
\epsscale{0.7}
\plotone{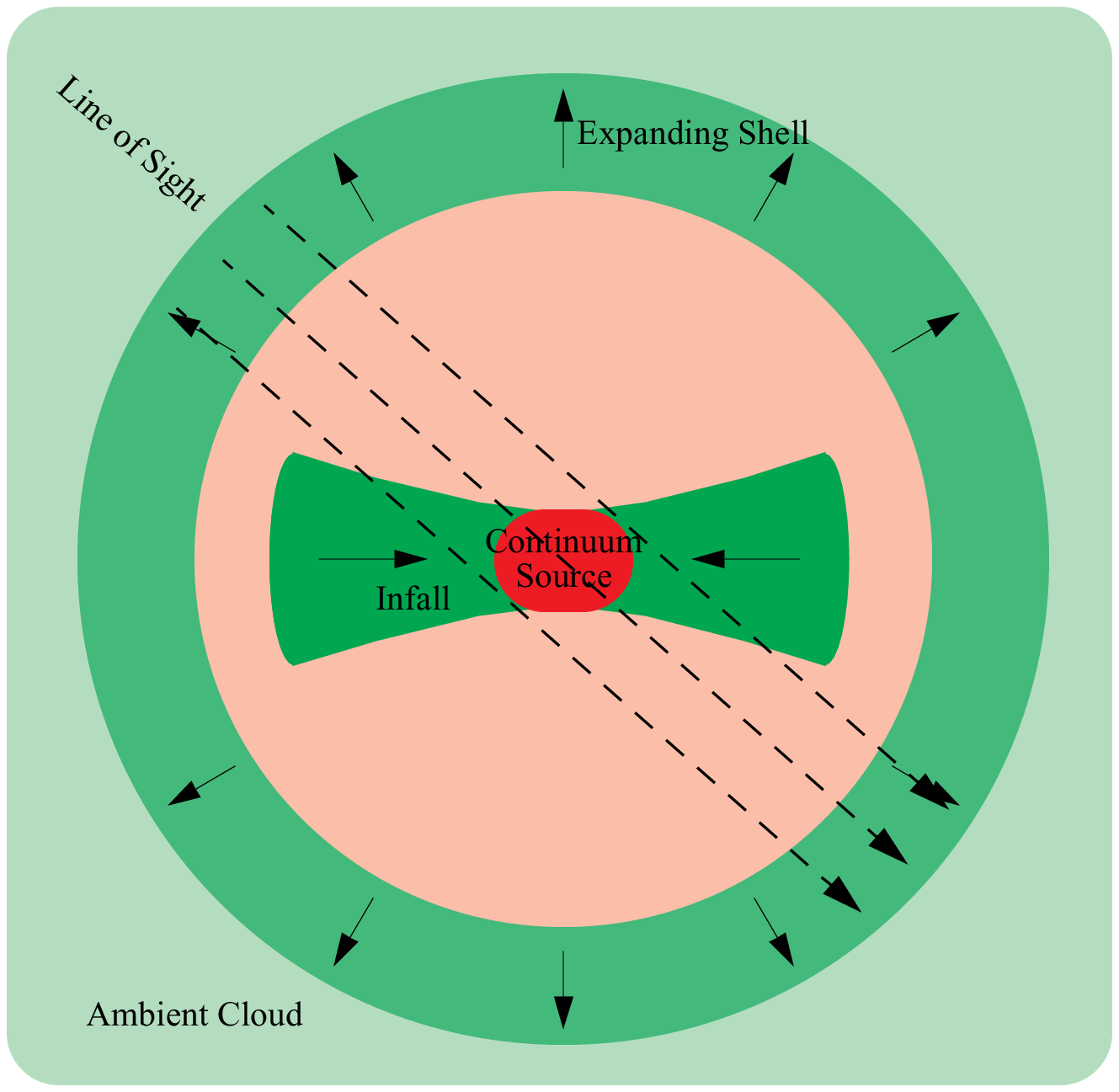}
\caption{A schematic drawing of the geometry discussed in Section
\ref{sec:discussion}. The figure shows a possible line of sight from
which the proposed geometry matches the observations. Solid arrows
represent motions in the molecular gas, while dotted-line arrows
represent lines of sight from the observer through the continuum
source and absorbing molecular material.}
\label{fig:cartoon}
\end{figure}

\begin{figure}
\plotone{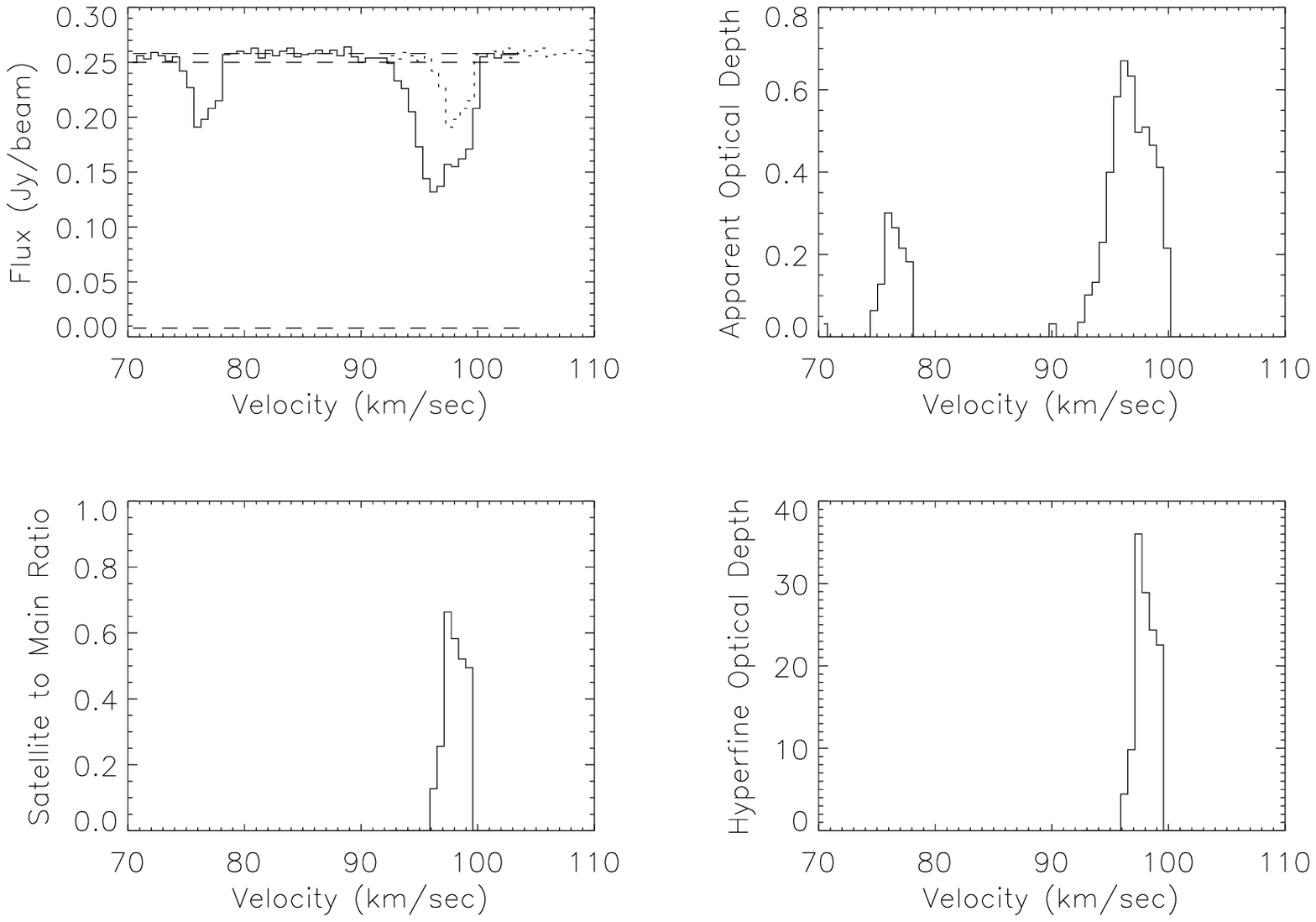}
\caption{A sample BnA array spectrum of \nhthree\ from G28.20-0.05,
with the calculated satellite to main hyperfine component ratio, and
the apparent and hyperfine optical depths. The first panel shows the
actual spectrum. The dashed lines show the continuum level at 0.258
\jyperbeam, the continuum level minus $2\sigma$ at 0.250 \jyperbeam,
and the lower detection limit at $2\sigma$, 0.008 \jyperbeam. The
dotted line shows the same spectrum shifted by 1.71~MHz, which is the
difference in rest frequency between the main hyperfine component and
the inner satellite hyperfine component. The second panel shows the
apparent optical depth. The third panel shows the line ratio
calculated for those channels in which both hyperfine components are
detected. The fourth panel shows the resulting hyperfine optical depth
as calculated for the main hyperfine component.}
\label{fig:samplespectra}
\end{figure}

\begin{deluxetable}{ccccccc}

\tablecolumns{6}
\tablecaption{The Data}
\tablehead{\colhead{Date} & \colhead{Line} & Energy\tablenotemark{a} (K) & \colhead{Array} & \colhead{Beam ($''$)} & \colhead{$\Delta v$ (\kms)} & \colhead{RMS (\mjyperbeam)}}
\startdata
Feb 14, 2003 & (1,1) & 23.4 & D & $4.0 \times 2.6$ & 0.6 & 5.2 \\
Feb 14, 2003 & (2,2) & 64.9 & D & $4.0 \times 2.6$ & 0.6 & 5.2 \\
May 27, 2003 & (3,3) & 125 & D & $3.1 \times 2.4$ & 0.3 & 3.4 \\
Oct 9, 2003 & (2,2) & 64.9 & BnA & $0.34 \times 0.23$ & 0.6 & 4.1 \\
Oct 10, 2003 & (3,3) & 125 & BnA & $0.39 \times 0.30$ & 0.6 & 3.9 \\

\label{tab:data}
\enddata

\tablenotetext{a}{\citet{ho83}}

\end{deluxetable}

\end{document}